\documentclass{nature_wplot}

\usepackage{gensymb}
\usepackage{graphicx}
\usepackage[final]{pdfpages}
\title{Absorptive pinhole collimators for ballistic Dirac fermions in graphene}

\author{*Arthur W. Barnard$^{1}$, Alex Hughes$^{1}$, Aaron L. Sharpe$^2$, Kenji Watanabe$^3$, Takashi Taniguchi$^3$ \& *David Goldhaber-Gordon$^2$}

\begin{document}

\maketitle

\begin{affiliations}
 \item Department of Physics, Stanford University, Stanford, California 94305, USA
 \item Department of Applied Physics, Stanford University, Stanford, California 94305, USA
 \item National Institute for Materials Science, 1-1 Namiki, Tsukuba 305-0044, Japan
\end{affiliations}

\begin{abstract}
Ballistic electrons in solids can have mean free paths far larger than the smallest features patterned by lithography. This has allowed development and study of solid-state electron-optical devices such as beam splitters\cite{a1,a2} and quantum point contacts,\cite{a3,a4} which have informed our understanding of electron flow and interactions. Recently, high-mobility graphene has emerged as an ideal two-dimensional semimetal that hosts unique chiral electron-optical effects due to its honeycomb crystalline lattice.\cite{a5,a6,a7,a8,a9,a10,a11,a12,a13,a14} However, this chiral transport prevents simple use of electrostatic gates\cite{a15} to define electron-optical devices in graphene. Here, we present a method of creating highly-collimated electron beams in graphene based on collinear pairs of slits, with absorptive sidewalls between the slits. By this method, we achieve beams with angular width 18\degree or narrower, and transmission matching semiclassical predictions. 
\end{abstract}

In the absence of scattering, electrons propagate freely as coherent waves, analogous to light in free space. Capitalizing on this behaviour, electron-optical elements including beam splitters,\cite{a1,a2} quantum point contacts (QPCs),\cite{a3,a15} lenses,\cite{a16} wave guides,\cite{a17,a18} and mirrors\cite{a4} have been fashioned in solid state two dimensional electron systems\cite{a19} (2DESs). Recently, encapsulation of graphene in hexagonal boron nitride (hBN)\cite{a20,a21} has enabled novel manifestations of refractive ballistic transport including quasiparticle dynamics in superlattices\cite{a13}, snake states,\cite{a22} Veselago lenses\cite{a12} and beam splitters.\cite{a23} While electrons in conventional semiconductor 2DESs can be collimated by QPCs, a collimated electron source in graphene remains conspicuously missing. Such a source could play a key role in diverse applications: ballistic transistors,\cite{a9,a24} flying qubits,\cite{a25} and electron interferometers.\cite{a26} However, theoretical proposals\cite{a27,a28} have yet to be realized and no robust demonstration of collimation has been reported to date. 

Here, we demonstrate experimentally and validate computationally an electron collimator based on a collinear pair of pinhole slits in hBN-encapsulated graphene. We show that grounded edge contacts\cite{a20} – analogous to peripheral surfaces painted black in an optical system – can efficiently remove stray electron trajectories that do not directly traverse the two pinholes, leaving a geometrically defined collimated beams. An absorptive pinhole collimator is constructed from an etched graphene heterostructure with a two-chamber geometry wherein independent electrodes make ohmic contact to each chamber (Fig. 1a). The contact to the bottom chamber (red, Fig. 1a) serves as the source for charge carriers while the contact to the top chamber (black, Fig. 1a) acts as an absorptive filter. To realize a collimating configuration, the filter contact (F) is grounded and the source contact (S) is current-biased; charge carriers are isotropically injected from the source, but only those trajectories that pass through both pinhole apertures reach the graphene bulk. Applying a uniform magnetic field can steer the collimated beam. For an uncollimated configuration, the filter and source contacts are electrically shorted. 

Our device consists of hBN-encapsulated graphene etched into a hall-bar-like geometry with the voltage probes replaced by collimating contacts (Fig. 1b). The hBN layers are both $d_{\rm{BN}} \sim 80\ nm$ thick and the device is assembled on $d_{\rm{ox}} = 300 nm$ $SiO_2$ atop a degenerately doped silicon substrate used as a back gate to tune charge carrier density n. To test the collimation behavior of an individual injector in the ballistic regime, we perform a nonlocal magnetotransport measurement, injecting from one collimator and probing trajectories that reach across the width of the device ($W_{\rm{dev}}=2\ \mu m$) in the collimated and uncollimated configurations (green, blue respectively, Fig. 1c). We inject from the lower right collimator (labeled S4,F4) throughout this Letter and in this case, measure the voltage of the upper right collimator (labeled S3,F3) relative to a reference (F1). In the presence of a B-field, electron trajectories that pass from the injector to collector flow from the injector at an angle $\theta=\sin^{-1}\frac{qBW_{\rm{dev}}}{2\hbar\sqrt{n\pi}} $, where $q$ is the quasiparticle charge. From this, we find that the angular full width at half maximum (FWHM) is $70\degree$ when injecting in the uncollimated configuration and 18⁰ when injecting in the collimated configuration.

For an uncollimated source\cite{a3}, the angular conductance is expected to go as $G(\theta)=\frac{2 e^2}{h}\sqrt{\frac{n}{\pi}}w_0\cos(\theta)$ where $\frac{2 e^2}{h}\sqrt{\frac{n}{\pi}}$ is the flux density at the Fermi level, and $w_0\cos(\theta)$ is the projected width of the contact. The collector has an acceptance angle of $\frac{w_0}{W_{\rm{dev}}}\cos(\theta)$ , leading to an expected $\cos^2(\theta)$ distribution ($\theta_{FWHM}=90\degree$). The $70\degree$ FWHM for our uncollimated data is in reasonable agreement with this expectation given that the reference contact collects more electrons at higher B-fields and thus suppresses the signal at high angles. 

In our collimators, the flux density at the Fermi level is identical to that in a single slit, but the projected width is geometrically defined by the pinhole width $w_0$ and pinhole separation $L_0$. For small angles $|\theta|<\tan^{-1}w_0/L_0$, the projected width $w(\theta)=\cos(\theta)\left[w_0-L_0|\tan(\theta)|\right]$ (left, Fig. 1d). At larger angles, no carriers should transmit, yielding:
\begin{equation}
G(\theta)=\frac{2 e^2}{h}\sqrt{\frac{n}{\pi}}w_0\cos(\theta)\left[w_0-L_0|\tan(\theta)|\right];\ \ \ |\theta|<\tan^{-1}\frac{w_0}{L_0}.
\end{equation}
Convolving over the acceptance angle of the collector (see Supplementary Information for details), we calculate the angular conductance distribution (middle, Fig. 1d) for both the uncollimated case (blue) and the collimated case (green) with $w_0=300\ nm$ and $L_0  = 850\ nm$, consistent with the fabricated collimator dimensions. The FWHM of the collimator emission is $22\degree$ for theory and $18\degree$ for experiment (right, Fig. 1d); showing that our injectors efficiently filter wide-angle trajectories and transmit narrowly collimated beams. 

Having established that the angular distribution of injected charge carriers is well-described by semiclassical theory, we now measure our collimators’ conductance to determine how efficiently electrons traverse the pinholes. For this, we bias the injector in the collimating configuration (F4 grounded) and measure the current reaching all remaining electrodes as a function of gate voltage (Fig. 2a). The conductance of the collimator tunes sublinearly with n: $(G\sim\sqrt{n-n_0})$ (dotted line, Fig 2a). This qualitatively agrees with semiclassical expectations $(G\sim\sqrt{n})$: integrating equation (1) over all angles, we expect 
\begin{equation}
G=\frac{4e^2}{h}\sqrt{n/\pi}\left[\sqrt{L_0^2+w_0^2}-L_0\right]
\end{equation}
The small offset $n_0\sim 1.6\times 10^{11}\  cm^{-2}$ in our measurement appears to result from diffraction by collimator slits (Fig. 2a, see Supplementary Information for details). Comparing equation (2), with the fit in Fig. 2a (neglecting $n_0$), indicates a conductance that is 35\% of expectations. This is a lower bound for the transmission probability, because the collimating filter (F4) can reabsorb electrons that have diffusely scattered off of device edges.

To understand the impact of diffuse scattering and better estimate the transmission probability, we measure the current collected at specific detectors as a function of B-field. Having sourced $I_{\rm{source}}=50\ nA$, we collect current in detectors collinear with (red, blue, Fig. 2b) and adjacent to (black, Fig. 2b) the injector. Current collected at the collinear detector with a wide acceptance angle (red) peaks near $B=0$ since the collimated beam travels straight across the device. The apparent background current is ~3-5\% of $I_{\rm{source}}$. At $B\sim120\ mT$, ballistic cyclotron orbits instead reach the adjacent detector, leading to a prominent peak in current detected at S1 (black) with F1 grounded. Coincident with this peak, the diffuse background of the collinear detector dips since ballistic trajectories are consumed by the adjacent detector, reducing the number of electrons that eventually find their way into the collinear detector. 
 
In light of the nontrivial diffuse background, we measure current with a narrow acceptance angle at the collector, rejecting most scattered electrons and thus better determining the transmission probability of the collimator. The resulting doubly-collimated beam (blue) has a FWHM of $8.5\degree$. Together all these collinear apertures act as a single collimator with $L_0=3,750\ nm$ (the separation between the farthest-apart apertures). All of the injected current passes through the first aperture, so the fractional current collected should be $
\frac{G(w_0=300 nm,\ L_0=3,750 nm)}{(G(w_0=300 nm,\ L_0=0))}=0.040$. The maximum of the doubly-collimated peak is 0.056 (Fig. 2b). Subtracting a background of 0.005-0.015 (see Supplementary Information for details), suggests transmission through the full path is $1.18\pm0.12$ times the expected value. The 20\% beamwidth-narrowing observed above for a single collimator ($18\degree$ vs $22\degree$ expected) may indicate modest focusing, which would be consistent with slightly-enhanced transmission through the double collimator. The excellent quantitative agreement shows that charge carriers transmit nearly perfectly from slit to slit. By demonstrating not only narrow beams but also high transmission probabilities, our measurements show that absorptive pin-hole filtering could produce low-noise, coherent, collimated beams of electrons in 2DESs that cannot be depleted by electrostatic gating.

Having experimentally demonstrated that absorptive pinhole collimators can controllably emit electron beams in hBN-encapsulated graphene heterostructures, we illustrate our technology’s utility by aiming a beam at the edges of our graphene device to learn about the low-energy scattering behavior of etched edges in these heterostructures. We perform three simultaneous nonlocal resistance measurements (Fig. 3a) to probe the specularity of reflections off various edges of the device. In Fig. 3b, we map $R_1=\frac{V_1}{I_{\rm{in}}}$  as a function of B-field and electron density. In both the electron-doped and hole-doped regimes, a peak near $B=0$ corresponds to ballistic quasiparticles being collected by the collinear contact in the absence of magnetic deflection. Peaks in $R_1$  also appear at higher fields, primarily in the hole-doped regime (n<0). For reference, we plot contours corresponding to cyclotron radius $r=W/2$. Any features outside the parabolas $(r<W/2)$ cannot correspond to direct ballistic quasiparticle transport across the width of the device, and must involve scattering. These data imply that holes undergo multiple reflections at high B-fields, suggesting that the edges may scatter more specularly when hole-doped than when electron-doped. 

To directly probe the specularity of reflections in our device, we perform a collimated transverse-electron focusing (TEF) measurement.\cite{a4,a29} Probe $V3$ at the lower left detector is even more sensitive than traditional TEF measurements to scattering that modifies ballistic trajectories, since here the injector and detector have narrow emission and acceptance angles respectively. $R_3=\frac{V_3}{I_{\rm{in}}}$  as a function of electron density and B-field has several distinct features associated with specific cyclotron radii (Fig. 3c), particularly for hole doping. At $r_1=1.25 \mu m$ there is a sharp peak with a FWHM of $\sim 300\ nm$ in both the hole and electron regimes. Though a conventional TEF peak would occur at $r_1=L_{\rm{lat}}/2$ where $L_{\rm{lat}}=2.3\ \mu m$ is the lateral separation of injector and detector, our measured peak corresponds to slightly greater cyclotron radius. This is expected for our collimator geometry: we illustrate the expected $r_1$ trajectory in Fig. 3a and plot its corresponding contour in Fig. 3c, indicating excellent agreement with our measurement (see also Supplementary Information for calculation). Trajectories at $r_1$ are insensitive to edge scattering, while at smaller r (larger B), additional peaks imply specular reflection. In the electron-doped regime, the presence of a prominent peak at $r_1$ with no appreciable secondary peak suggests completely diffuse scattering, while in the hole-doped regime, the presence of a significant secondary peak suggests appreciable specular reflection.

To validate this understanding and quantitatively determine the degree of specularity, we next carry out semiclassical, device-scale simulations. Modeling the fabricated device geometry including all ohmic contacts, we simulate electron emission from the injector, allowing for reflection off edges, and interaction with floating or grounded ohmics. With two free parameters, transmission of ohmics $p_{\rm{trans}}$ and probability of diffuse edge scattering $p_{\rm{diffuse}}$, we simulate the measurement configuration shown in Fig. 3a-c (see Supplementary Information for simulation details). The striking similarities between simulation ($p_{\rm{trans}}=67\%$ and $p_{\rm{diffuse}}=100\%$) and measurement suggest that edge scattering is diffuse in our device in the electron-doped regime (Fig. 3d, see Supplementary Video 1 for visualizing a B-sweep). Similar analysis yields $p_{\rm{trans}}=10\%$ and $p_{\rm{diffuse}}=67\%$ in the hole-doped regime, quantitatively demonstrating significant electron-hole asymmetry in both ohmic contact properties and specularity of edge scattering in our device

The strong agreement between semiclassical theory and experiment for both individual collimators and our entire collimating device indicates that absorptive collimation in high-mobility graphene devices can be predictably and robustly applied in a variety of geometries, opening the door for scientific and technological use of narrow electron beams in 2DESs. For example, Klein tunneling\cite{a7,a8} and Andreev reflections\cite{a30} are highly angularly-dependent phenomena whose experimental signatures are obscured in typical transport experiments. In such cases, collimation-based measurements will illuminate the physics by quantitatively testing transmission and reflection at specific angles rather than integrated over a range of angles as in past experiments. In addition, novel technologies such as ballistic magnetometers may be built on the sharp magnetotransport features we achieve. Collimated sources are an important addition to the growing toolbox of electron-optical elements in ballistic graphene devices that enable a new class of transport measurements.

\begin{methods}
\subsection{Sample Fabrication}
Flakes of graphene (from highly oriented pyrolytic graphite, Momentive Performance Materials ZYA grade) and of hBN (from single crystals grown by high-pressure synthesis as in Ref. 21) were prepared by exfoliation (3M Scotch 600 Transparent Tape) under ambient conditions (35-60\% relative humidity) on n-doped silicon wafers with 90 nm thermal oxide (WRS materials). The heterostructure was assembled by a top-down dry pick-up technique as described in (20). The completed heterostructure was deposited on a chip of n++-doped silicon with 300 nm thermal oxide (WRS materials). Polymer residue from the transfer process was removed by annealing the sample in a tube furnace for 1 h at 500 C under continuous flow of oxygen (50 sccm) and argon (500 sccm). Device patterns were defined by e-beam lithography and reactive ion etching as described in (13). Ohmic contacts were established to the device using electron-beam evaporated Cr/Au electrodes to the exposed graphene edge as described in (20). 

\subsection{Measurement}

All measurements were performed at 1.6K in the vapor space of a He flow cryostat with a superconducting magnet. Lock-ins (Stanford Research Systems SR830) at 17.76 Hz were used in all measurements; voltages were measured with Stanford Research Systems SR 560 voltage preamplifiers and currents were measured with Ithaco 1211 current preamplifiers. The charge density n was calculated from Shubnikov-de-Haas oscillations $\left(\frac{n}{V_{\rm{g}}} = 5.51\times 10^{10}\ cm^{-2} V^{-1}\right)$, in good agreement with the expected geometric capacitance.

\end{methods}

%% Put the bibliography here, most people will use BiBTeX in
%% which case the environment below should be replaced with
%% the \bibliography{} command.

% \begin{thebibliography}{30}
% \end{thebibliography}

\bibliographystyle{naturemag}
\bibliography{bibtex_for_collimator_arxiv}

%% Here is the endmatter stuff: Supplementary Info, etc.
%% Use \item's to separate, default label is "Acknowledgements"

\begin{addendum}
 \item We thank M. Lee and T. Petach for fruitful discussions. This work was financially supported by the Gordon and Betty Moore Foundation through Grant GBMF3429, by a Nano- and Quantum Science and Engineering Postdoctoral Fellowship (A.B.), by a Ford Foundation Predoctoral Fellowship (A.S.) and a National Science Foundation Graduate Research Fellowship (A.S.). K.W. and T.T. acknowledge support from the Elemental Strategy Initiative conducted by the MEXT (Japan). T.T. acknowledges support from JSPS Grant-in-Aid for Scientific Research under grants 262480621 and 25106006. Part of this work was performed at the Stanford Nano Shared Facilities (SNSF).
 \item[Author Contributions] A.B., D.G.-G., A.H., and A.S. conceived of the measurements. A.S. fabricated the device. A.B., A.H., and A.S. performed transport measurements. A.B. and A.H. performed numerical simulations. A.B. wrote the manuscript with input from all other authors. K.W. and T.T. grew the bulk hBN crystals.
 \item[Competing Interests] The authors declare that they have no
competing financial interests.
 \item[Correspondence] Correspondence and requests for materials
should be addressed to \\ A.B. ~(email: barnarda@stanford.edu) or D.G.-G. ~(email: goldhaber-gordon@stanford.edu).
\end{addendum}

\begin{figure}
\includegraphics[width=400pt]{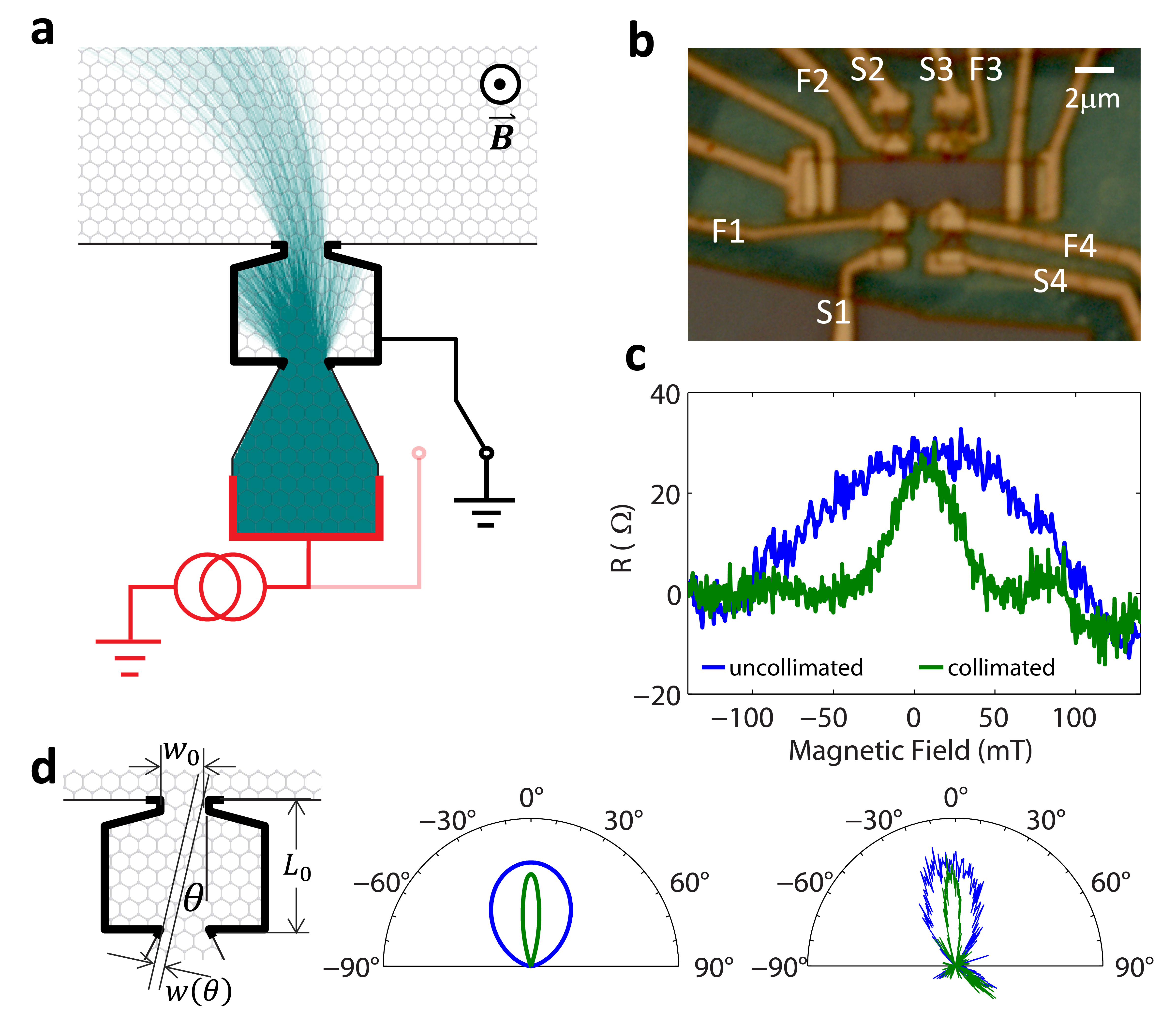}
\centering
\spacing{1}
\caption{\textbf{Absorptive pinhole collimators. a.} Double pinhole collimator schematic. Current is sourced from bottom contact (red), passes through the bottom aperture and is either absorbed by the top contact (black) or passes into the device bulk. Only trajectories that pass through both apertures reach the bulk, producing a collimated beam. The collimated beam is steered by an external B-field. \textbf{b.} Optical micrograph of device with four collimators in a hall-bar-like geometry. \textbf{c.} Measuring angular distribution. Nonlocal resistance at $n=1.65\times 10^{12} cm^{-2}$ is plotted with $V_{S3F3}$ measured relative to VF1 when current is sourced from both S4 and F4 (blue) and only from S4 while F4 is grounded (green). The narrowness of the central peak for the F4-grounded data results from collimation. \textbf{d.} Theoretical collimation behaviour vs. experiment. Left: diagram of effective collimator width $w(\theta)$ at a fixed angle for semiclassical trajectories. Middle: polar plot of theoretical angular dependance for a 300 nm wide point contact (blue) and a $w_0=300\ nm$, $L_0=850\ nm$ collimator (green). right: experimental data from \textbf{(c)} mapped to angle. }
\end{figure}

\begin{figure}
\includegraphics[width=400pt]{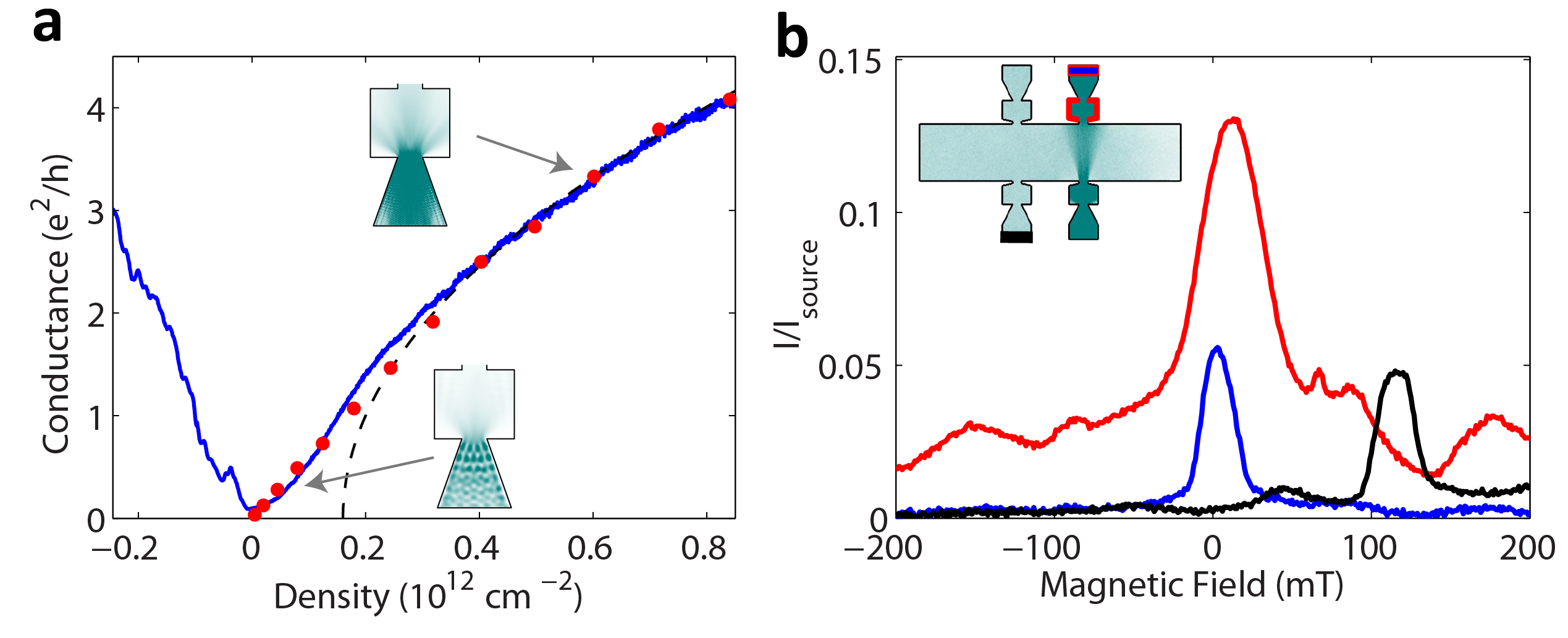}
\centering
\spacing{1}
\caption{\textbf{Conductance of single and paired collimators. a.} Conductance of collimator measured in a three terminal configuration: S4 is current-biased, F4 is grounded, and all remaining terminals are measured with a single current amplifier. The measured conductance (blue) scales as $\sqrt{n-n_0}$ (black dotted line), qualitatively agreeing with semiclassical ballistic conduction of bulk graphene. Numerical solutions to the 2D Dirac equation (red dots) account well for low density effects associated with diffraction. \textbf{b.} Conductance measurements through angularly sensitive collectors. Current is collected at F3+S3 (red), S3 (blue) and S1 (black) with all remaining contacts grounded. F3+S3 has a broad background due to diffuse edge scattering and imperfect ohmic contacts. S3 has a FWHM of 8.5 degrees due to double-collimation and has minimal diffuse background. The peak height of S3 indicates nearly perfect ballistic transmission.  }
\end{figure}

\begin{figure}
\includegraphics[width=400pt]{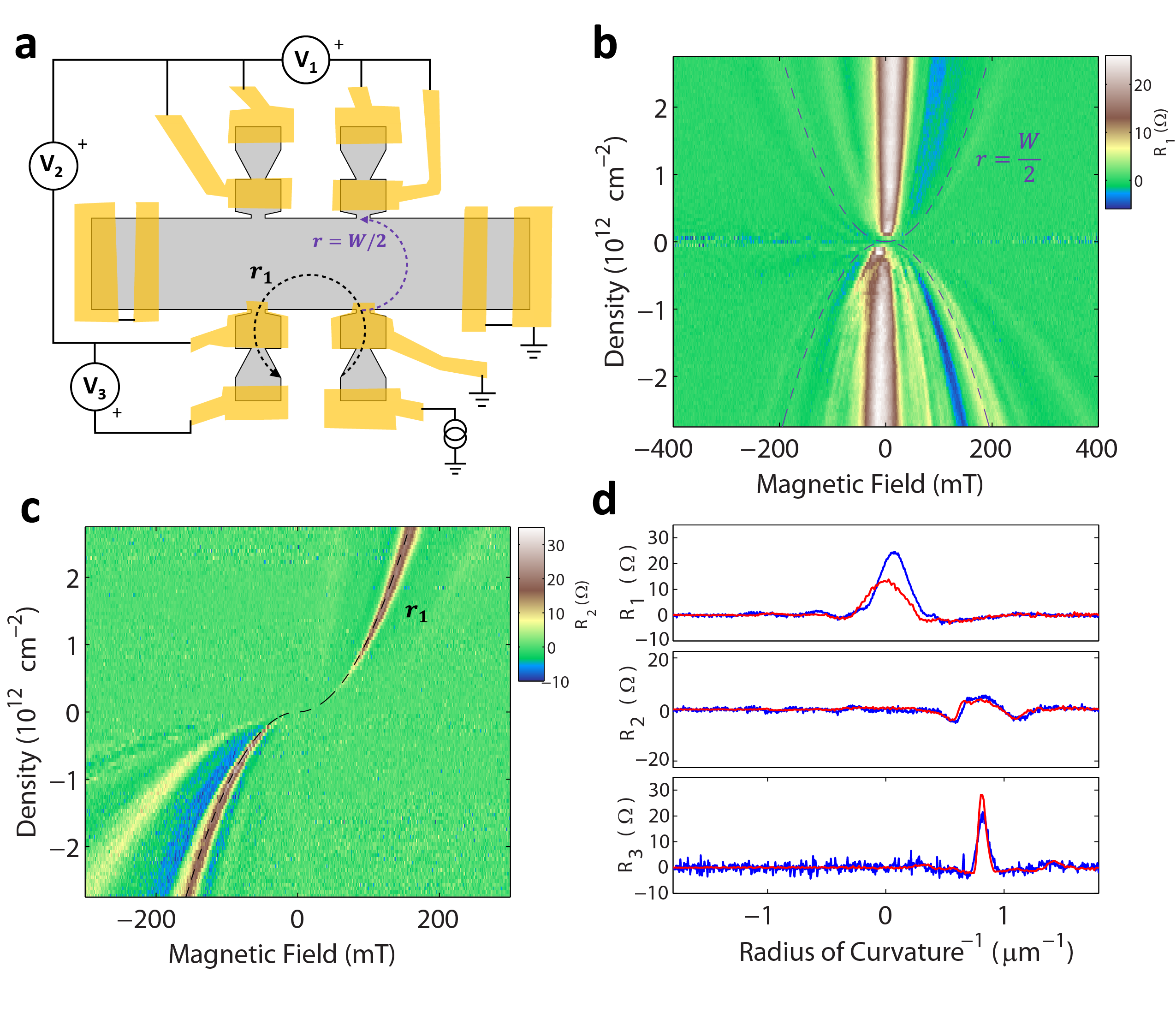}
\centering
\spacing{1}
\caption{\textbf{Probing edge scattering. a.} Nonlocal resistance measurement schematic. \textbf{b.} Resistance map characterizing angular profile of injected trajectories. A central peak near $B=0$ corresponds to the beam passing straight across the width of the device (a small angular offset is due to fabrication imperfections). The remainder of the electron doped regime $(n> 0)$ is nearly featureless, while the hole doped regime $(n < 0)$ has several auxiliary peaks. Dotted lines correspond to cyclotron orbits with radius equal to the half of device width $(r=W_{\rm{dev}}/2)$; features outside the two parabolas cannot correspond to direct ballistic trajectories between injector and collector. \textbf{c.} Collimated transverse electron focusing. A sharp feature at $r_1  = 1.25\ \mu m$ corresponds to trajectories that pass through four pinholes. Features at higher magnetic field must involve specular reflections off of the device edge. There is no such feature on the electron side, while there is a noticeable band on the hole side. \textbf{d.} Comparison of experimental data to semiclassical simulation. Experimental data (blue) are taken at $n=2.7\times 10^{12}\ cm^{-2}$, and simulation (red) assumes fully diffuse edge scattering and 67\% ohmic transmission.}
\end{figure}
\newpage



\section*{Supplementary material}

# Supplementary Information: Absorptive pin-hole collimators for ballistic Dirac fermions in graphene

Arthur W. Barnard, Alex Hughes, Aaron L. Sharpe,
Kenji Watanabe, Takashi Taniguchi, and David Goldhaber-Gordon

## 1 Semiclassical conductance calculation

Similar to the calculation of current through 1D channels or quantum point contacts, our type of collimator can be treated as a finite constriction connecting two electron reservoirs. Net current only flows in an energy band for which there is a Fermi energy mismatch $\Delta E = E_{f_2} - E_{f_1}$, where $E_{f_1}$ and $E_{f_2}$ are the Fermi energies of each reservoir. In two dimensions, the angular flux density is:

$$\frac{dJ}{d\theta} = \frac{e}{\hbar} \frac{d^2 n}{d\theta dk} \Delta E \qquad (1)$$

where $n$ is the electron density, $e$ is the charge of an electron, $k$ is the electron wave vector magnitude, and $\theta$ is the angle of electron propagation. At the Fermi level,

$$\frac{d^2 n}{d\theta dk} = \frac{g_k k_f}{4\pi^2} \qquad (2)$$

where $g_k$ is the degeneracy of a fixed $k$ state. In the case of graphene, the degeneracy is 4 due to the valley and spin degrees of freedom. This gives:

$$\frac{dJ}{d\theta} = \frac{2e^2 k_f}{\pi h} V_{\text{bias}} = \frac{2e^2}{h} \sqrt{\frac{n}{\pi}} V_{\text{bias}} \qquad (3)$$

Where $V_{bias} = \Delta E/e$ is the voltage difference between reservoirs. The electron flux density is uniform and controlled by the 2D electron density. To calculate the angular current density $I(\theta)$, the width of the constriction at a given angle $w(\theta)$ must be known. For the case of the double pin-hole collimator, the projected width is $w(\theta) = \cos(\theta)[w_0 - L_0|\tan(\theta)|]$ in the range $|\theta| < \tan^{-1} w_0/L_0$ where $w_0$ is the width of an aperture and $L_0$ is the separation between apertures. This means that the current emitted over a narrow range of angles $\Delta\theta$ is:

$$I(\theta)\Delta\theta = V_{\text{bias}} \frac{2e^2}{h} \sqrt{\frac{n}{\pi}} \cos(\theta)[w_0 - L_0 |\tan(\theta)|] \Delta\theta. \qquad (4)$$

In our angular distribution measurement, we collect ballistic electrons across the width of the device $W_{\text{dev}}$ through a third pinhole collector as we vary the magnetic field. The collector aperture has an acceptance angle in the small angle approximation that goes as $\Delta\theta \approx \frac{w_0 \cos(\theta)}{W_{\text{dev}}}$. Convolving the injection angular distribution with the collector angular distribution results in the predicted nonlocal conductance:

$$G = \frac{2e^2}{h} \sqrt{\frac{n}{\pi}} \int_{\theta_B - \frac{w_0}{2W_{\text{dev}}} \cos(\theta_B)}^{\theta_B + \frac{w_0}{2W_{\text{dev}}} \cos(\theta_B)} d\theta \cos(\theta)[w_0 - L_0 |\tan(\theta)|] \qquad (5)$$

where $\theta_B = \sin^{-1}\left(\frac{eB}{h} \sqrt{\frac{\pi}{n}} W_{\text{dev}}\right)$ is the central angle injected that reaches across the width of the device at a given $B$. We integrated numerically supplementary equation (5) with $w_0 = 300$ nm

and $L_0 = \{0\,\text{nm}, 850\,\text{nm}\}$ to produce the plot in Fig. 1d in the main text. If the width of the collector is sufficiently small, supplementary equation (5) reduces to:

$$G(\theta_B) = \frac{2e^2}{h}\sqrt{\frac{n}{\pi}}\frac{w_0}{W_{\text{dev}}}\cos^2(\theta_B)\left[w_0 - L_0\left|\tan(\theta_B)\right|\right] \tag{6}$$

The full width at half maximum then comes from solving: $\cos^2\frac{\theta_{\text{FWHM}}}{2}\left[1 - \frac{L_0}{w_0}\tan\frac{\theta_{\text{FWHM}}}{2}\right] = \frac{1}{2}$. For narrow beams, this simplifies to: $\theta_{\text{FWHM}} = \frac{w_0}{L_0}$.

## 2 Transverse electron focusing

Unlike in traditional transverse electron focusing (TEF) measurements, the center of most cyclotron orbits exiting our double-pinhole collimators are *not* collinear with the device edge. Instead, the median center is displaced off the edge by $L_0/2$, and as a result, successive focusing peaks occur at irrational ratios. For an injector-collector separation $X_0$, the radii of curvature that result in peak conduction from the collimator to the adjacent collector follow the relationship:

$$r_n = \sqrt{\left(\frac{L_0}{2}\right)^2 + \left(\frac{X_0}{2n}\right)^2} \tag{7}$$

In our collimating device, $X_0 = 2.3\,\mu\text{m}$ and $L_0 = 850\,\text{nm}$. Two related consequences of the arc center offset are (1) the incident angle to edges is no longer normal ($\theta = \frac{\pi}{2}$) but rather $\theta_n = \tan^{-1}\frac{L_0}{nX_0}$ and (2) there is a fixed minimum radius $r_\infty = \frac{L_0}{2}$ below which no ballistic conduction should occur.

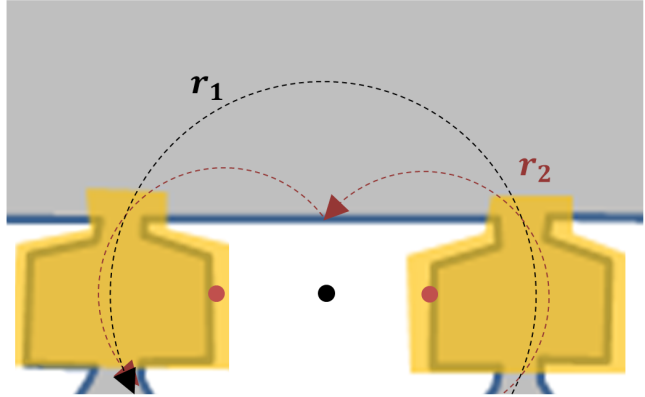

Supplementary Figure 1: diagram of TEF trajectories in our collimators. The center of each arc is plotted as a solid circle, and is offset from the edge by half the length of a collimator. The consequence of this: TEF peaks are not evenly spaced and the incident angle varies with $r_c$

## 3 Diffraction effects on transport: 2D Dirac equation simulations

Low energy electron excitations in graphene obey the 2D massless Dirac equation. This behavior implies that electrons in our collimators should be subject to diffraction, which may particularly affect electron transport at low electron densities. We thus perform finite-difference time-domain (FDTD) simulations of the Dirac equation in order to predict the transport behavior of our collimators.

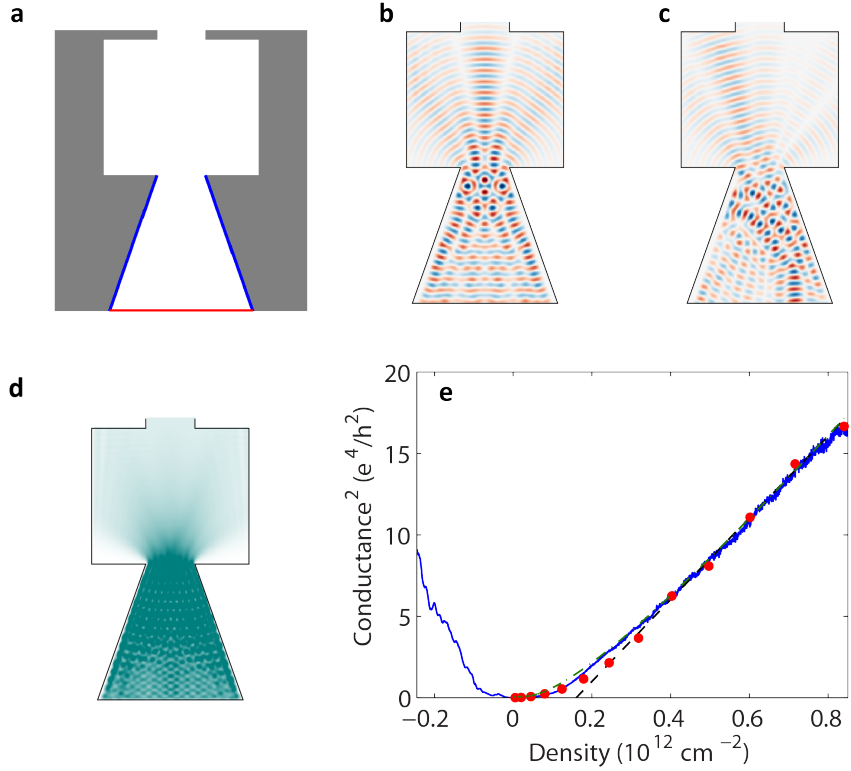

Supplementary Figure 2: a) Defined regions for simulation. gray corresponds to absorptive conditions, blue to reflective, and red to the injection region. b) real part of $u$ for a single plane-wave solution. c) particular instance of random electron injection. d) Magnitude of $J_y$ (the current density in the vertical direction). e) Same data as shown in Fig. 2a, with the y-axis squared to emphasize the square-root-dependence. Included is a phenomenological fit as a green dashed line

The Dirac equation for electrons in graphene is:

$$\pm i v_\text{f} \boldsymbol{\sigma} \cdot \nabla \Psi = -i\hbar \partial_t \Psi \tag{8}$$

where $\boldsymbol{\sigma}$ is a vector representation of the two Pauli matrices $\sigma_x$ and $\sigma_y$, $v_\text{f}$ is the Fermi velocity, $\partial_t$ connotes a single partial time derivative, and $\psi$ is the two-component single-particle wavefunction. Defining the two components as $\Psi \equiv \begin{pmatrix} u \\ v \end{pmatrix}$, and working in graphene's natural units ($\hbar = 1$ and $v_\text{f} = 1$), supplementary equation (8) can be expressed as two coupled equations:

$$\partial_t u \pm \partial_x v \mp i \partial_y v = 0 \tag{9}$$

$$\partial_t v \pm \partial_x u \pm i \partial_y u = 0 \tag{10}$$

We discretize supplementary equation (9) and (10) using a staggered space and staggered time approach [1], and apply reflective or absorptive boundary conditions. Reflective boundary conditions are employed by setting $u$ to zero at boundaries and allowing $v$ to propagate freely [2]. Absorptive boundaries are employed using the "absorbing potential" approach [1]. In order to inject electrons, we couple in a time-varying complex potential at the edge of the collimator ohmic contacts (Supplementary Fig. 2a). We inject plane-waves (Supplementary Fig. 2b) at several injection angles to be summed over later.

In order to calculate total conductance of the collimator, we iteratively sum the plane-waves solutions with random amplitudes to approximate integration over all spatial probability amplitudes. During each iteration (Supplementary Fig. 2c) we compute the current density:

$$\boldsymbol{J} = ev_{\text{f}} \Psi^{\dagger} \boldsymbol{\sigma} \Psi \tag{11}$$

and then average $\boldsymbol{J}$ over all iterations (Supplementary Fig. 2d). The conductance plotted in Fig. 2a (and Supplementary Fig. 2e) then results from integrating over $\boldsymbol{J}$ at the top aperture.

These simulations give an absolute prediction for the collimator conductance. However, as discussed in the main text, the measurement in Fig. 2a is partially attenuated by electron reflections and reabsorption by the filter ohmic. Thus, we allow one free parameter: the absolute conductance. Good agreement between simulation and experiment is evident from the proper scaling of the conductance with density: namely, in the limit of large $n$, the conductance goes as $G \sim \sqrt{n - n_0}$. $n_0$ is not a free parameter of the fit, yet it agrees well with the data.

This effect is likely due to the reflective boundary conditions of the first pinhole, as well as the subsequent diffraction off the slit. One way to phenomenologically parameterize these effects is to assert that the effective width of the apertures is reduced by a length proportional to the Fermi wavelength $\lambda_{\text{f}}$. We find that making the mapping $w_0 \to w_0 - \lambda_{\text{f}}$ in equation (2) of the main text reproduces our data well. With this ansatz, we fit our data with one free scaling parameter (Supplementary Fig. 2e) and find excellent agreement. The generality of this exact mapping for different collimator geometries bears further investigation, however the basic observation that finite-wavelength effects moderately reduce the conductance relative to semiclassical theory seems robust.

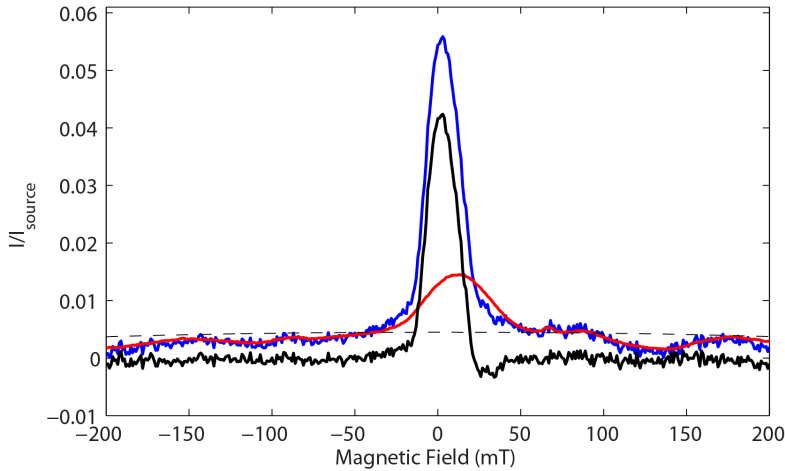

Supplementary Figure 3: Diffuse background subtraction for measuring transmission probability. Normalized current S3 is plotted in blue. The normalize current S3C3 is scaled by a factor of 9 to best fit the diffuse background of S3 (red line). The difference of the two, S3-(S3C3)/9, is plotted as solid black. This represents a lower-bound for the background-subtracted conductance. Alternatively, a fit assuming a completely isotropic background is plotted as a dotted black line. This should be the minimum background signal

## 4 Background subtraction for determining transmission probability

In order to accurately measure the transmission probability of electron trajectories through our collimators, we measured a doubly-collimated signal. We found that this significantly reduces the impact of a diffuse background, however, a small background remains even in this configuration.

Here, we discuss our approach regarding background subtraction. There is insufficient information to fully disentangle the contributions of ballistic transport and the diffuse background, so we instead compute upper and lower bounds for the background.

For the upper-bound, we first observe that the wide-collection angle signal (S3C3) has a nearly identical functional form for the diffuse background as for the narrow collection angle signal (S3). We plot S3C3/9 over S3 in Supplementary Fig. 3, and observe significant agreement between the two curves at $|B| > 50$ mT. If we assume that the majority of the background current in S3 results from electrons scattering inside the collimator chamber, then the background is simply S3C3/9. We plot S3C3/9-S3 (solid black line, Supplementary Fig. 3) which has a peak height of 0.042.

Alternatively, if we assume that the contacts in C3 absorb all incoming electrons, then the only available background electrons must pass ballistically through both collimator apertures. For simplicity, if we assume an isotropic background, the background should be proportional to $\cos^2 \theta$, where $\theta = \sin^{-1}\left(\frac{eB}{\hbar}\sqrt{\frac{\pi}{n}}L_0\right)$. We fit such a curve to the regime $|B| > 50$ mT (dotted black line, Supplementary Fig. 3). Using this background, the conductance peak height is 0.051.

The true diffusive background is likely between these two limits given that the contacts have reasonably high (but not perfect) transmission probabilities.

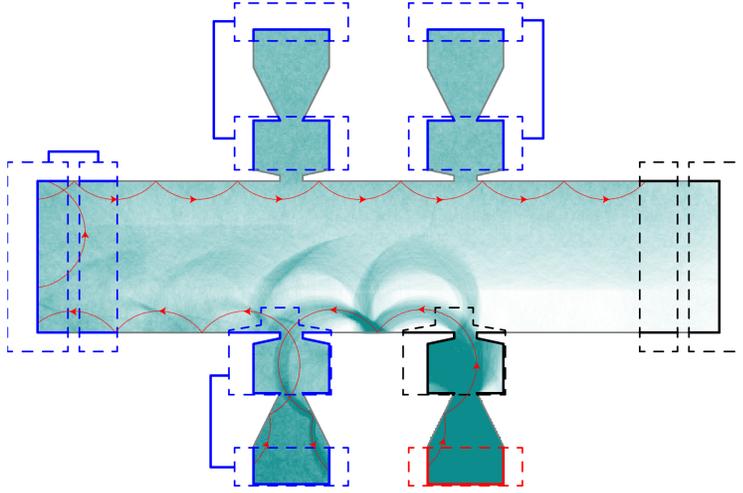

Supplementary Figure 4: A simulation of our hallbar at the first magnetic focusing peak ($r_c = 720$ nm) with $p_{\text{scatter}} = 0$ and $p_{\text{trans}} = 0.667$. Carriers are sourced at the red edge contact and flow through the device until they reach one of the grounded (black) contacts. Blue contacts are floating, and all dashed lines indicate physical contacts. One particular carrier's path is highlighted in red with arrows pointing in the direction of travel. Darker background color indicates higher density of trajectories.

## 5 Ballistic simulations

Because most carriers injected into our device scatter several times off sheet edges before being absorbed by a grounded ohmic, it is important to consider all possible trajectories to accurately model our device. Charge carriers are first injected into the graphene sheet at a random position along the source contact and at a random angle. These carriers follow their semiclassical trajectories until they hit an edge or they scatter in the bulk with a characteristic scattering length $l_{\text{scatter}}$. When they hit an edge, the appropriate behavior–scatter, specularly reflect, refract, transmit–is determined from the nature of the edge hit and from control parameters $p_{\text{scatter}}$, the probability an edge will randomly scatter, and $p_{\text{trans}}$, the probability that an ohmic will transmit.

This process is repeated for each injected carrier until it is absorbed in a grounded contact. In the simulations both here and discussed in the main text, 40,000 carriers were injected into the sheet.

For current measurements, current drains are grounded and the number of carriers absorbed into each contact is recorded. The measured current flowing out a particular contact is given by $I_{\text{contact}}/I_{\text{source}} = n_{\text{contact}}/N$ where $n_{\text{contact}}$ is the number of carriers that end up in a particular grounded contact and $N$ is the total number of injected carriers. In order to ensure detailed-balance, floating voltage leads are simulated by absorbing incident carriers, followed by re-emitting the same carriers at random positions and angles. The voltage in these leads should be proportional to the flux of carriers through the edge contacts. Thus, the voltage is given by $V_{\text{meas}} \propto \phi_{\text{contact}}/L_{\text{contact}}$ where $\phi_{\text{contact}}$ is the number flux of carriers through the contact and $L_{\text{contact}}$ is given by the length of the contact along the graphene sheet.

To model edge behavior in our device, we need to account for the complete fabricated device geometry. In particular, certain ohmics were mildly misaligned, resulting in asymmetric edge contact to the device region in C1 and C4. Based directly off of optical micrographs of the device (e.g. Fig. 1b), we defined the geometry and contacts as shown in Supplementary Fig. 4. The graphene sheet is defined first (grey), then contacts are added on top (dashed lines) in the measurement configuration from Fig. 3 of the main text. Shown is a snapshot at a constant magnetic field corresponding to a radius of curvature of $r_c = 720$ nm. We have highlighted one trajectory as a guide to the eye. Note that this trajectory is absorbed and re-emitted several times by floating ohmics.

The scattering probability of the edges is an experimental unknown that we try to understand based on comparison with our simulations. We thus conducted magneto-transport simulations at several values of $p_{\text{scatter}}$, which is the probability that a given charge carrier will scatter following a cosine distribution normal to the edge, while holding $p_{\text{trans}} = 0.67$ constant. We compare the area under the first (no bounce) and second (one bounce) TEF peaks, and as seen in Supplementary Fig. 5, we simulated magnetic field sweeps over a range from $p_{\text{scatter}} = 0$ to $p_{\text{scatter}} = 1$. The second peak's area decreases linearly with increasing scattering probability while the first remains largely unchanged, so we compare the ratio with our experimental data both on the hole and electron side to find that the scattering probability on the electron side is $p_{\text{scatter}} \approx 1$, and $p_{\text{scatter}} \approx 0.5$ on the hole side.

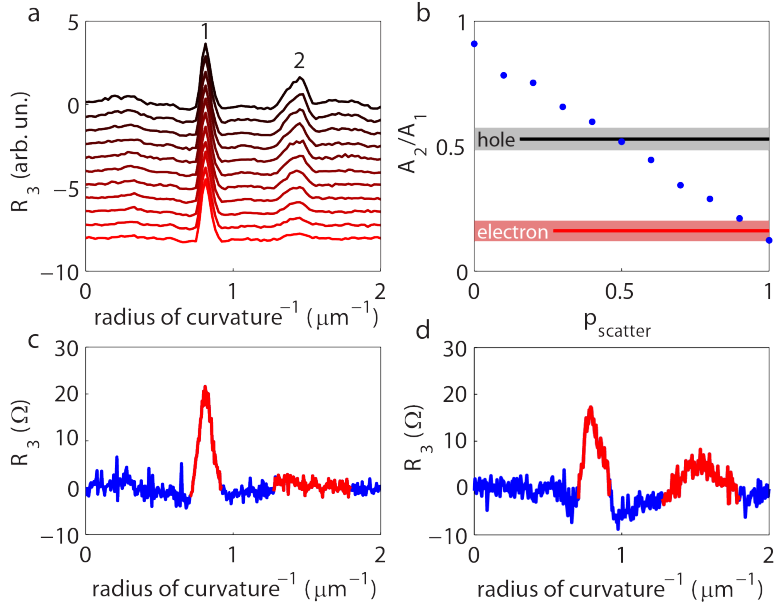

Supplementary Figure 5: Edge scattering simulations. **a**. Simulated magnetic field sweeps using the contact scheme shown in Supplementary Fig. 4. The edge scattering probabilities range from $p_{\text{scatter}} = 0$ (red) to $p_{\text{scatter}} = 1$ (black). There is little change in the first focusing peak (1), but the second peak (2) decreases in height and area with increasing scattering probability. **b**. Ratio of area under the second magnetic focusing peak to the area under the first decreases linearly with increasing edge scattering probability. The hole- and electron-side ratios for our data are plotted in black and red, respectively. The lighter-colored shading indicates the uncertainty in the peak ratios from our data. Measurement data for electron doped regime (**c**) and hole doped regime (**d**). integration ranges for peaks are highlighted in red.

To optimize $p_{\text{trans}}$, we observe the functional form of all three non-local resistances and find the best-fit conditions. For the electron side, this corresponds to $p_{\text{trans}} = 0.67$. For the hole side, the contact resistances are higher, and the best fit simulation is $p_{\text{scatter}} = 0.67$ and $p_{\text{trans}} = 0.1$. As is evident, the scattering probability is modestly higher than the above analysis, resulting from a weak dependence of $A_2/A_1$ on $p_{\text{trans}}$. We plot the final fit on the hole side (analogous to Fig. 3d in the main text) in Supplementary Fig. 6. The fit is qualitatively good, though not as striking as for the electron side. Given that there is substantially greater spurious emission due to high contact resistance as well as higher specularity of reflections for the hole-doped data, it is actually surprising to have as close a fit as observed.

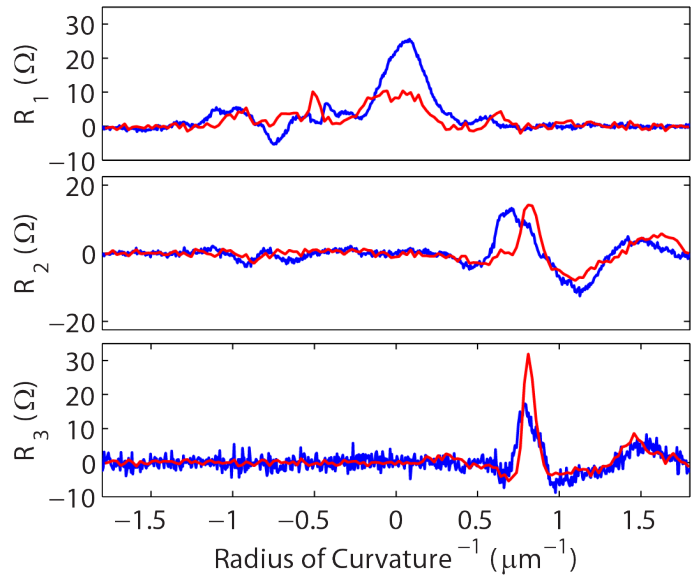

Supplementary Figure 6: Simulation vs experiment in hole doped regime. Data is plotted in blue, and simulation results ($p_{\text{scatter}} = 0.67$ and $p_{\text{trans}} = 0.1$) are plotted in red.


\end{document}